\documentclass{ws-mpla}

\begin{document}

\markboth{A. Gardini and P. M. Ricker}
{Simulations of Hot Bubbles in the ICM}

%
\catchline{}{}{}{}{}
%

\title{SIMULATIONS OF HOT BUBBLES IN THE ICM}

\author{\footnotesize ALESSANDRO GARDINI}

\address{Department of Astronomy, University of Illinois at Urbana-Champaign, 1002 W Green Street\\
Urbana, IL, 61801, USA\\
gardini@astro.uiuc.edu}

\author{PAUL M. RICKER}

\address{Department of Astronomy, University of Illinois at Urbana-Champaign, 1002 W Green Street\\
Urbana, IL, 61801, USA\\
pmricker@astro.uiuc.edu}

\maketitle

\pub{Received (Day Month Year)}{Revised (Day Month Year)}

\begin{abstract}
We review the general properties of the intracluster medium (ICM)
in clusters that host a cooling flow,
and in particular the effects on the ICM of the injection 
of hot plasma by a powerful active galactic nucleus (AGN).
It is observed that, in some cases, the hot plasma produces
cavities in the ICM that finally detach and rise, perhaps buoyantly.
The gas dynamics induced by the rising bubbles can help explain 
the absence of a cooled gas component in clusters with a cooling flow.
This scenario is explored using numerical simulations. 

\end{abstract}

\ccode{PACS Nos.: 98.65.Hb}

\section{Clusters of Galaxies and the ICM}

Clusters of galaxies were first identified as regions 
in the universe containing overdensities of galaxies.
For example, Abell\cite{Abell58} defined a cluster as containing
at least 50 bright galaxies within a circular area of sky of radius
1.5~Mpc/$h$.\footnote{$h$ is Hubble's constant in units of
100\ km\ s$^{-1}$\ Mpc$^{-1}$.}
It was recognized as early as 1930 that clusters must contain more
matter than is present in their component galaxies if they are to
remain gravitationally bound.\cite{Zwicky33}
In fact, galaxies contribute only a few percent of the mass of a cluster,
which ranges from about $10^{13} M_\odot$ for galaxy groups to about
$10^{15} M_\odot$ for the richest clusters.
About 15-20\% of the mass exists in the form of a hot, diffuse
plasma known as the intracluster medium (ICM), first detected in
the early 1970s.\cite{Meekins71,Gursky71}
Due to its high temperature ($\sim 10^7 - 10^8$~K),
the ICM produces most of its
radiation in the form of thermal X-rays whose surface brightness is
strongly concentrated toward the center of a cluster.
Measurements of the galaxy velocity dispersion\cite{Zwicky37}, the
temperature of the ICM\cite{Mitchell77,Mushotzsky78}, 
and more recently gravitational lensing\cite{Fort94} 
show that at least 80\% of the mass of clusters exists
in some nonluminous form (dark matter) which is most likely
nonbaryonic.\cite{Cavaliere98}

Galaxy clusters are the largest gravitationally bound objects in the
universe. According to the hierarchical scenario of structure formation,
they grow through infall and merging of smaller cluster and groups.
This process appears to have continued to the present day.\cite{Gunn72}
Morphologically we can classify clusters as being regular, having
single brightness peaks and ellipsoidal or spherical shapes,
or irregular, having multiple brightness peaks or irregular shapes.
Irregular clusters appear to be undergoing or have recently
undergone merger events.
Regular clusters may have gone some time since their last mergers.
However, since the timescale to re-establish equilibrium 
after a major merger event is of the order of several
Gyr\cite{Ricker01,Ritchie02}, and since projection effects can hide
evidence for mergers, the fraction of clusters that appear regular
most likely provides an underestimate of the fraction that have
recently undergone mergers.
Still, to first order hydrostatic and virial equilibrium is a
reasonable approximation for many clusters; hence cluster properties
are often computed under the assumption of spherical symmetry.

Typical X-ray luminosities of clusters
range between about $10^{44}$ and $10^{45}$~erg~s$^{-1}$.
The X-ray emission of the ICM is due to thermal bremsstrahlung
for $T > 3 \cdot 10^7 K$ 
while line cooling becomes very important at lower temperatures.
While to first approximation the ICM appears to be isothermal,
there is evidence that the temperature profile decreases at large 
radii\cite{DeGrandi02}.
Many clusters that host a cooling flow also show 
a temperature drop in the central region.
Because of its high temperature, the ICM is almost completely ionized.
The surface brightness follows the gas distribution: the gas
density shows a central core and decreases at larger radii.
The ICM is also rarefied; the electron density $n_e$ ranges
from some $10^{-2}$ cm$^{-3}$ in the cores and then decreases below
$10^{-4}$ cm$^{-3}$ at the limits of detectability.
The gas metallicity is a fraction of solar ($\sim 0.3 Z_\odot$)
and is in almost solar proportion.
This implies that the gas is not completely primordial but rather 
has been partially processed in galaxies 
and then stripped by ram pressure or reinjected into the medium by
supernova explosions. A metallicity excess is detected in
the cores of cooling-flow clusters\cite{DeGrandi01}.
Measurements of Faraday rotation in sources beyond clusters also show 
that the ICM hosts a magnetic field of about 0.1 to 10 $\mu G$\cite{Govoni01}.

\section{The Cooling Flow Problem}
\label{Sec: cooling flows}

Due to its X-ray emission the ICM loses energy over time. 
We can define the cooling timescale as
$t_{\rm cool} = - (d\, \ln T / dt)^{-1}$; 
for gas emitting via thermal bremsstrahlung this yields
$$t_{\rm cool} = 8.5 \times 10^{10} \Big{(}{n_p\over {10^{-3} {\rm\ cm}^{-3}}}\Big{)}^{-1}
             \Big{(}{T\over 10^8 {\rm\ K}}\Big{)}^{1/2} {\rm\ yr}
$$
if the gas cools isobarically\cite{Sarazin88}.
Because of its low density, the ICM is thus expected to shine 
for billion of years without appreciable cooling, except
in the very central regions of some clusters. 
Here the gas density is large enough that
the cooling time becomes shorter than the Hubble time.
For these clusters we can define a cooling radius $r_{\rm cool}$
within which $t_{\rm cool}$ is less than the age of the universe.
The cooling radius can be larger than $\sim 100$~kpc.

In this paper we will refer to those clusters with an
appreciable $r_{\rm cool}$ as cooling-flow clusters,
because in such clusters a cooling flow should be established
in the absence of sources of heat\cite{Fabian94}.
The classic picture of a cooling flow consists of a subsonic
flow of cooling gas toward the cluster center.
In a single-phase medium, the cooling gas should lose energy
and be compressed by the pressure of the overlying gas, thus
cooling still faster and eventually becoming invisible in X-rays
altogether\cite{Fabian84}.
Cooling flow clusters generally do have surface brightness
distributions that are more concentrated than those of clusters
without cooling flows\cite{Peres98}, but
homogeneous cooling models produce surface brightness profiles
that are much steeper than observed.
Inhomogeneous cooling can mitigate this problem\cite{Nulsen86,Johnstone92};
in this picture, cold clouds condense out of the flow over a wide
range of radii, allowing some of the gas near the center to
remain hot.

The temperature profiles of cooling-flow clusters often show
a marked drop in the central regions, as expected.
Because mergers increase the entropy of the ICM, the presence
of a cool core is often taken to imply that a cluster is relaxed,
even for those clusters that host 
a cooling flow together with substructures (e.g., Abell 85).
A giant elliptical or a cD galaxy sits
at the center of every cooling-flow cluster\cite{Eilek04,Mathews03}.

Recent observations have forced major revisions in our
understanding of cooling flows.
For the observed cores, we expect to see gas temperatures 
as low as $10^6$~K (0.1 keV) and mass deposition rates as
high as 1000 $M_\odot$~yr$^{-1}$.
However, high-resolution
XMM/Newton\cite{Peterson01,Tamura01,Kaastra04},
ASCA\cite{Makishima01}, and Chandra\cite{David01} spectra
of cool cores do not show the expected signatures 
of cooling below 1-2 keV, namely
the emission lines of metals that dominate
cooling at such temperatures.
Moreover, observations in the optical\cite{Kent79,Crawford99}, 
radio\cite{Edge01}, and infrared\cite{Edge02} show much
smaller amounts of cooled gas than predicted by earlier X-ray
estimates of the mass accretion rate\cite{Peres98}. 
The observed star formation rates\cite{Crawford99,Mittaz01} 
are also smaller than expected from cooling flows.
Some mechanisms have been suggested\cite{Fabian01,Peterson03}
that could suppress the cooling 
flux at low energies and yet maintain the large cooling rates,
but besides requiring some fine tuning, they do not address the question
of the repository for most of the cold gas.
So the other possible conclusion is that at most a small
fraction of the ICM cools below 1-2 keV\cite{Molendi01}. 

Maintaining gas at keV temperatures for a period several times longer than
its cooling time requires one or more heating mechanisms. 
Any successful mechanism has to satisfy several requirements 
to match the observations consistently.
First, it has to provide sufficient heating to balance the cooling losses.
This heating must be properly tuned: too little heat will not impede
cooling, while too much heat will produce an outflow from the core. 
This suggests that the heating process should be self-regulated.
For example, the mass deposition may trigger the heating process and
in turn be quenched by it.
The heating mechanism must also affect (perhaps indirectly)
the entire mass deposition region in order to prevent partial cooling
in the less well-heated regions.
The heating mechanism also must preserve the entropy profile of the ICM,
which decreases toward the cluster center.
This rules out a purely convective mechanism, 
which would act if the entropy were decreasing with radius.
Finally, the heating should not destroy the metallicity profile, which
shows a marked rise in the core\cite{Bohringer02}.

Candidate heating mechanisms include 
thermal conduction by electrons in the
ICM\cite{Narayan01,Zakamska03,Voigt04}, 
reconnection of magnetic fields\cite{Soker90},
turbulent mixing\cite{Kim03},
and injection of hot plasma by the active galactic nuclei
(AGN) hosted by the central galaxies of many clusters. 
In this paper we will focus on this last possibility.

\section{The Interaction of Active Galaxies with the ICM}

Recent observations show that almost all cooling-flow clusters 
host radio sources in their centers\cite{Eilek04}.
Often these sources are rather faint, and it is not clear
if a correlation exists between their radio power and the strength
of the cooling flow\cite{Voigt04,Kaastra04,Birzan04}.
Anyway, about 71\% of the cDs in cooling-flow clusters 
are radio loud compared to only 23\% of non-cooling-flow cluster 
cDs\cite{Burns90,Ball93}, and this suggests a connection 
between the AGN activity and the presence of the cooling flow, 
at least in clusters that host a cD.

In active radio sources, an AGN drives strong outflows in the form of jets
which are highly collimated and composed of relativistic particles. 
The jet production process is not yet completely understood, but it is 
generally assumed to involve some magnetohydrodynamic mechanism
which acts on the material falling 
onto a supermassive black hole\cite{Celotti01}.
The plasma inside the jets produces radio synchrotron and
synchrotron self-Compton emission and is sometimes observed
in the optical and in X-rays.
The jets lose energy by doing work against the surrounding medium,
finally coming into pressure equilibrium with the ICM. 
At this point the jets inflate lobes which are
filled with relativistic plasma and magnetic field.
This plasma still emits synchrotron radiation,
so the lobes are visible in radio waves.

The shapes of radio sources are determined by the interaction of the
jets with their surroundings. For radio sources not connected
with a cooling flow, the jets travel straight for tens
or hundreds of kpc before ending in a hot spot or undergoing a sudden
transition and continuing as broad tails.
The central radio sources in cooling-flow clusters, on the other hand,
show mostly a disturbed morphology: even when they host a radio-loud core,
collimated jets exist only on kpc scales or below. After this
the energy flow continues in a less collimated manner into the ICM, 
and the lobes take the shape of large bubbles.

Shortly, the radio galaxies 
should evolve through three stages\cite{Heinz98,Begelman03,Begelman04}. 
At the beginning the radio plasma is driven by the momentum of the jets and
inflates cigar-shape cocoons which are overpressured with the respect 
to the ambient medium. 
Hence the cocoons expand laterally meanwhile the jets are 
lenghtening their channels, and the supersonic lateral expansion 
become rapidly comparable to the lenghtening.
However, the cocoon expansion relents then by the way, 
and after the expansion velocity has became subsonic
the evolution of the cocoon is governed by bouyancy.

A rather different scenario suggests that the observed ultrarelativistic jets
are instead embedded in an unobserved subrelativistic flow\cite{Blandford99}.
In this case the outflow from the AGN would carry a significant amount
of momentum, and the cocoons would rise supersonically pushed by the jets.

High angular resolution X-ray images obtained by Chandra show 
in some cases
depressions of the X-ray flux, suggestively referred to as cavities,
that are coincident with AGN radio lobes.
Cavities are observed, for example, in Perseus\cite{Fabian00}, 
Hydra A\cite{McNamara00}, Abell 2052\cite{Blanton01}, 
and Centaurus\cite{Sanders02}.
They are usually some tens of kpc across and a similar distance 
away from the cluster center.
They are often surrounded by bright rims of denser and cooler gas
which presumably has been displaced, uplifted or entrained by the
hot plasma during the formation of the cavity or its subsequent motion.

In some cases weak shocks are observed\cite{Nulsen04} but no strong shocks, 
thus is assumed that the radio-emitting plasma in these cavities is 
almost in pressure equilibrium with the surrounding ICM.
As stated above, it is then easy to show that the lobes 
filled by relativistic particles and magnetic field 
should have a smaller specific weight than the ICM and
act as bubbles that finally detach and rise buoyantly.
In simulations the cavities rise at about 1/3 the sound speed in the ICM
\cite{Churazov01}.

Chandra X-ray images also show cavities that are not coincident with
bright radio lobes, for example in Abell 2597\cite{McNamara01}, 
Perseus\cite{Fabian00}, and Abell 4059\cite{Heinz02}. 
These structures are referred as ghost cavities. 
The relativistic electrons in the radio lobes are expected 
to lose enough energy via synchrotron emission to become invisible
in the radio after 50 to 100~Myr. 
If this interpretation is correct,
the ghost cavities are buoyantly rising relics of a radio
outburst that ended at least 50 to 100~Myr ago\cite{Soker02}.

Cavities, both filled and ghost, have been clearly detected in 15 clusters 
that host an AGN in a cool core. 
A correlation has been shown to exist between the mechanical
luminosity of the cavities 
$L_{\rm mech}$ and the power of the radio source.
Here $L_{\rm mech}$ is defined as $pV/t_c$, where $p$ is the pressure
inside the cavity, $V$ is the volume of the cavity, and $t_c$ is its age.
A correlation also exists between $L_{\rm mech}$ and the
X-ray luminosity due to the cooling ICM gas.
These correlations suggest that the cavities are indeed powered
by AGN and that the AGN activity level is related to
the accretion of cooling gas.
However, the mechanical luminosity is not large enough for rising bubbles
to balance radiative cooling by themselves except in a few cases\cite{Birzan04}.

The commonly observed cold fronts discovered by Chandra also provide
hints regarding the effect of AGN on the intracluster medium.
Cold fronts are sharp discontinuities in
X-ray surface brightness marking boundaries between hotter and
colder masses of gas in the ICM. The density and temperature jumps
inferred for cold fronts through deprojection analysis are consistent
with continuous pressure changes across the fronts, showing that
the discontinuities are not shocks\cite{Markevitch00}.
While cold fronts are usually explained as remnants of past merger
events, they have been observed also in apparently regular clusters
such as RXJ~1720.1+2638\cite{Mazzotta01}, Abell 1795\cite{Markevitch01}, 
2A~0335+096\cite{Mazzotta03}, and MS~1455.0+2232\cite{Mazzotta01bis}.
The gas sloshing induced by the motion of a pre-existing cavity 
may explain these observations\cite{Mazzotta03}.

Finally, radio-emitting regions as large as 1~Mpc are sometimes
found in the outskirts of clusters far from radio sources that
might power them\cite{Rottgering97}.
Some of these `radio relics' are quite far from cluster
centers\cite{Giovannini00}.
These relics may trace the shock waves generated by merger
events\cite{Roettiger99,Miniati01} .
Alternatively, they may represent very late-stage ghost cavities 
that have been re-energized at some point\cite{Ensslin01,Hoeft04}.

\section{AGN as a Source of Heat}

As shown above, recent X-ray observations show that the cores
of cooling-flow clusters are often far from being settled down;
rather, central radio sources drive gas motions that should heat
the intracluster medium.
It is tempting to argue that this heating addresses the problems
with the cooling flow model described in Section~\ref{Sec: cooling flows}.
However, this cannot be the whole story, and in any case many
details remain to be worked out.

Bubbles might heat the ICM in any of several different ways.
Part of the jet plasma energy is transferred to the ICM during the process
of bubble formation\cite{Heinz98},
and also work can be done on the ICM by the adiabatic expansion
of the bubbles as they rise\cite{Begelman04}.
If the bubbles expand subsonically (on average), 
both of these processes transfer energy into the ICM as weak shocks
and sound waves, thus affecting the ICM at a distance
from the bubbles themselves and possibly outside 
the cool core\cite{Ruszkowski04}. 

The rising of a bubble by buoyancy is a process driven 
by the potential energy released as the surrounding medium falls in 
around the bubble to fill the space it occupied.
It is easy to show\cite{Churazov02,Birzan04} 
that the potential energy dissipated as the
bubble rises is given by the variation of the entalphy 
$H = \gamma p V / (\gamma -1) $ of the bubble.
This energy is mostly deposed in the wake of the bubble as kinetic and
potential energy of the entrained gas, as turbulence and possibly as 
gravity waves, and finally converted to heat by damping and/or mixing.
This process is called sometimes {\it effervescent heating}
\cite{Begelman01,Ruszkowski02,Begelman04}.

Another possibly effective heating 
mechanism\cite{Quilis01,Churazov02,DallaVecchia04} consists
of the mixing of the cold gas uplifted and/or entrained from the cluster center
with the hot gas from more external regions that slides off the bubbles during
their rise. This mechanism would be as more effective in heating the cool
core as more material is radially displaced 
without disrupting the metallicity profile.

It can be observed that the cold gas uplifted by the bubbles 
could expand adiabatically into the surrounding warmer ICM 
due to its higher pressure, thus cooling further before it mixes
\cite{Bohringer02,McCarthy03}.
This would explain the X-ray observation that the shells of material
surrounding cavities appear sometimes colder than the ICM in the center
of the cluster.\cite{Schmidt02,Blanton03}

If the bubbles survive long enough, they reach a hydrostatic
equilibrium position where they expand only laterally, again helping
to distribute energy globally in the cooling-flow region.
However, the final fate of the bubbles is unclear. 
While the bubble plasma should finally mix with the ICM, 
the persistence of ghost cavities suggests that it does 
not mix significantly on timescales comparable to the bubble rise time.
This is a rather difficult behaviour to reproduce even in numerical
simulations unless we invoke the presence of a confining magnetic 
field\cite{Robinson04}.
It has been also observed in simulations, 
that bubbles driven by subrelativistic flows
tend to resist disruption longer\cite{Omma04}.

Two major points that AGN heating models must address
are the lack of correlation between the power of the central radio source
and the strenght of the cooling flow and the fact that the cavities 
are not a universal phenomenon.
Radio sources are not even detected in some cooling-flow clusters,
and cavities are reported only in a fraction of those clusters 
that host a radio source.
A possible solution is then to assume a duty cycle: the cooling
of the gas triggers the AGN activity, and the resulting ejected plasma 
inflates the radio lobes that in turn reheat the ICM and stop cooling.
The timescale of the radio lobes' evolution should then be a few times
shorter than the cooling timescale\cite{Quilis01,Voit01,Ruszkowski02}.

Even though it is often observed that the bolometric luminosity 
of the AGN does not account for the mechanical power needed
to balance the cooling\cite{Churazov02}, the details of the
bubble heating mechanism can be important in determining the AGN
duty cycle.

Indeed, simulations have been performed to explore the effect of 
recurrent bursts of activity from the AGN,
but these have shown that the mixing of the ICM that follows the rising 
of a bubble can endure even for 1 Gyr, thus effectively reducing the 
cooling that led to the activity in the first place\cite{Basson03}.

While it appears that the rising of the bubbles and related phenomena
do not alter the entropy profiles, the observed metallicity gradients
place constraints on the effeciency with which bubble can mix 
the ICM\cite{Bohringer04}.

Other questions that deserve a better understanding include
the possibility that other heating mechanisms cooperate 
with AGN to prevent cooling, the role of AGN in generating cold fronts, 
the interaction of bubbles with pre-existing bubbles or with galaxies,
and the revitalization of ghost cavities by mergers.

\section{The Numerical Approach}

Exploring the dynamics of the hot plasma in the cluster environment
is a very complex problem, owing to its intrisinc three-dimensionality,
and to the chaotic interaction of the hot plasma with the ICM.
The main tool consists of hydrodynamic simulations, although
some analytic\cite{Soker02,Mathews03} and semi-analytic\cite{Kaiser03}
models exist.
We review here some of the most recent results.

Br\"uggen \& Kaiser (2002)\cite{Bruggen02a} 
perform a pair of very high resolution simulations adopting 
a plane-parallel configuration and a twodimensional system of coordinates. 
They use the adaptive mesh FLASH\cite{Fryxell00} code to simulate 
a region of size 50 kpc $\times$ 100 kpc with a resolution of 25 pc.
They do not impose a jet or a bubble, but rather they inject energy 
locally into the ICM: this injection mimics the effect of the jet 
on the ICM and generates a buoyant cavity.
Although this is a rough approximation, this model is able
to show that the bubble motion has a significant effect on the ICM:
after the passage of the rising bubble, the average cooling time
in a gas layer increases by up to a factor 1.4 
with respect than the value set in the inital conditions. 

Br\"uggen (2003)\cite{Bruggen03} extends the previous work adopting
a more realistic density distribution; he assumes a modified 
Navarro, Frenk, \& White\cite{Navarro97} (NFW) profile
for both gas and dark matter and a bipolar distribution for the
energy injection.
Radiative cooling is also included \cite{Raymond76} 
and the ICM and the hot plasma respectively obey perfect-gas
equations of state with $\gamma=5/3$ and $\gamma=4/3$.
The simulation is performed again using FLASH 
and spans a region of size 140 kpc $\times$ 140 kpc 
centered on the cluster center, with a resolution of 68 pc;
the energy is injected at some distance from the center in two opposite
spherical regions of radius 0.5 kpc at the rate of 
$L=6\cdot10^{44}$ erg/s for 208 Myr. The simulation lasts $\sim$ 380 Myr.
This more realistic simulation confirms the previous results 
and also shows that cold gas from the cluster core is effectively 
uplifted by the rising bubbles and appears as bright rims 
in surface brightness maps.

Br\"uggen et al. (2002)\cite{Bruggen02b} 
describe a set of three three-dimensional simulations using 
the ZEUS-MP\cite{Stone92a,Stone92b} code. 
In the first two cases, they simulate a region of size
10 kpc $\times$ 10 kpc $\times$ 30 kpc with a resolution of 67 pc.
The assumed initial distributions of mass, electron density, temperature 
and pressure are those of the Virgo Cluster\cite{Nulsen95}.
Cooling is not included and $\gamma=5/3$ is used for both the gas phases.
Energy is injected in a sphere of radius $r=0.7$ kpc at a distance of
9 kpc from the cluster center.
Two different values of the energy injection rates are considered: 
$L=4.4\cdot 10^{41}$ erg/s and $L=3.8\cdot 10^{42}$ erg/s.
In a third case, the simulated region spans
16 kpc $\times$ 16 kpc $\times$ 20 kpc
while the source luminosity is $L= 10^{44}$ erg/s.
The most relevant result of these simulations consists in the
observation that cavities are quite difficult to detect, both in radio
and in X-rays. Thus it is suggested that a significant amount of
energy can be hidden in bubbles in clusters.

Basson \& Alexander (2003)\cite{Basson03} 
perform a set of two three-dimensional
simulations using the ZEUS-MP code adopting a spherical system of coordinates;
the grid spans $194 \times 64 \times 64$ cells in $r \times \theta \times \phi$
and is arranged so as to increase resolution along the polar axis, where
the jets are placed. The ICM is modeled by an isothermal $\beta-$profile
and the dark matter potential is derived accordingly. 
The computational domain is bounded by two spherical surfaces,
and the external radius appears to be $\sim 2800 \,kpc$
as derived from their pictures.
Cooling of the gas is accounted for, and the cooling function
is created composing expressions from the 
literature\cite{Sutherland93,Puy99,Tegmark97}.
Instead of injecting energy in spherical regions, two jets of plasma are
modeled in these simulations. 
These simulations show that a long-lived buoyancy-driven convective flow is 
established by the rising of the cavities. Even if the cluster reverts to
having a cooling flow, the convective motions are able to remove the cold gas
accumulating in the cluster core. Indeed there is a net outflow persisting
for timescales of about one order of magnitude longer than the time for
which the source is active.
 
Dalla Vecchia et al. (2004)\cite{DallaVecchia04}
perform a set of three-dimensional
simulations using the FLASH code in Cartesian coordinates.
They consider an isothermal cluster of mass $M=3\cdot10^{14} M_\odot$
and temperature $k_B T=3.1$ keV. The dark matter density follows an NFW profile,
so the gas density is derived accordingly\cite{Makino98}.
The computational domain covers a cubic region of 1.9 Mpc on side
with a resolution of 7.4 kpc.
Radiative cooling is considered\cite{Theuns02}. 
The energy is injected in spheres 
distributed in random directions in the cluster halo with a different energy 
in each simulation run. A duty cycle in the energy injection is imposed
and a new sphere is injected every 100 Myr, 
while the simulation lasts for 1.5 Gyr.
It is shown that a value of the injected energy exists
which effectively balances the energy loss due to cooling and 
preserves the ICM temperature. It is also shown that the cooling rate
is reduced not by the direct heating of the cooling gas, 
but by its convective transport to regions of lower pressure. 

Omma et al. (2004)\cite{Omma04} perform a set of two three-dimensional
simulations using the ENZO\cite{Bryan97,Bryan99} code, with an adaptive grid.
They simulate a computational box 635 kpc on side with a resolution
of 620 pc. Cooling is neglected, but the AGN outflow is modeled 
as a subrelativistic flow starting from the cluster center.
The cavities generated by this flow rise supersonically as long as the
source is active, driving a vortex of uplifted gas in their wake. 
As the AGN is switched off, the vortex region finally overtakes the cavity, 
filling it with cool plasma, which is overdense with respect 
to the surrounding ICM.
This overdensity is heated by the dissipation of the $g$ modes of the cluster
and is strongly excited by them before it falls back inwards.

Omma and Binney (2004)\cite{Omma04bis} explore the structural stability
of cooling flows adopting a set of five simulations similar to those
described above and varying the jets' parameters. 
They raise the possibility that the feedback mechanisms between cooling
gas and radio source breaks down in FRII systems, because they will
deposit the majority of the ejected energy outside the core.

Robinson et al. (2004)\cite{Robinson04}
perform a set of two-dimensional
simulations using the FLASH code with plane-parallel configuration.
The simulated region covers 120 kpc $\times$ 150 kpc with a resolution
of 29 pc.
They compare the evolution of bubbles with and without the presence
of conduction and of magnetic field. 
It is shown that the magnetic field is indeed 
necessary to preserve the bubble's integrity for a time long
enough to observe it as ghost cavity or for a few bubble rise times.

\section{Conclusions}

The new observations of the cores of cooling-flow clusters 
show clearly the interaction between the central radio source
and the ICM. However, the details of this interaction and
its relationship with the cooling flow problem are still uncertain.
The field needs new insights, both from the observational side and 
the theoretical side.

Deep X-ray observations will detect faint features due to shocks
and sound waves and perhaps discriminate between the various models
of the outflow from the AGN. Deep radio observations could probably
detect faint central radio sources or cavities, while observations
at different wavelengths could better determine the power of the sources
and test the correlation with the luminosity of the cooling flow.

Future simulations should consider several physical phenomena 
affecting the ICM and the ejected plasma
that have not yet been considered or properly modeled.
The role of magnetic fields deserves further examination,
particularly considering their effect on the thermal conductivity
of the ICM. The modeling of the AGN duty cycle and initial bubble
injection are still overly simplified. The equation of state of
the relativistic bubble plasma must be treated more carefully.
Finally, while the studies to date have considered bubbles in
idealized cluster atmospheres, clusters are known to undergo
mergers which dramatically affect their structure.
The turbulent mixing produced by these mergers and by the motion
of galaxies through the ICM should strongly affect the development
of AGN-blown bubbles, so it is important that the next generation
of bubble simulations consider clusters within their cosmological
context.

\section*{Acknowledgments}

AG and PMR acknowledge support under a Presidential Early
Career Award from the U.S. Department of Energy, Lawrence
Livermore National Laboratory (contract B532720) as well as
the National Center for Supercomputing Applications.
AG wishes to thank Pasquale Mazzotta, Silvano Molendi and Sabrina De Grandi 
for useful discussions; he also thanks Rossella Cerutti 
and Fabrizio Tavecchio for relevant reading suggestions.

\end{document}